\documentclass[twocolumn,english]{revtex4-1}
\usepackage[LGR,T1]{fontenc}
\usepackage[latin9]{inputenc}
\usepackage{units}
\usepackage{textcomp}
\usepackage{ifsym}
\usepackage{amssymb}
\usepackage{graphicx}
\usepackage{wasysym}
\usepackage{subscript}

\makeatletter

\DeclareRobustCommand{\greektext}{%
  \fontencoding{LGR}\selectfont\def\encodingdefault{LGR}}
\DeclareRobustCommand{\textgreek}[1]{\leavevmode{\greektext #1}}
\DeclareFontEncoding{LGR}{}{}
\DeclareTextSymbol{\~}{LGR}{126}

\@ifundefined{date}{}{\date{}}
\@ifundefined{showcaptionsetup}{}{%
 \PassOptionsToPackage{caption=false}{subfig}}
\usepackage{subfig}
\makeatother

\usepackage{babel}
\begin{document}

\title{{\Large{}Particle Ice Front Interaction - The Brownian Ratchet Model}}

\author{Michael Chasnitsky\textsuperscript{1{*}}, Victor Yashunsky\textsuperscript{2}
and Ido Braslavsky\textsuperscript{1}}

\affiliation{\textsuperscript{1}The Robert H Smith Faculty of Agriculture, Food
and Environment, The Hebrew University of Jerusalem, Rehovot, Israel}

\affiliation{\textsuperscript{2}Laboratoire PhysicoChimie Curie, Institut Curie,
PSL Research University - Sorbonne Universit\textasciiacute es, UPMC-CNRS
- Equipe labellis\textasciiacute ee Ligue Contre le Cancer; 75005,
Paris, France.}

\email{michael.chasnitsky@mail.huji.ac.il}

\selectlanguage{english}%

\date{December 2017}
\begin{abstract}
We treat the problem of particle pushing by growing ice as a free
diffusion near a wall that moves with discrete steps. When the particle
diffuse away from the surface the surface can grow, blocking the particle
from going back. Elementary calculations of the model reproduce established
results for the critical velocity \emph{v\textsubscript{c}} for particle
engulfment: \emph{v\textsubscript{c}\textasciitilde{}}1\emph{/r}
for large particles and \emph{v\textsubscript{c}\textasciitilde{}Const}
for small particles, \emph{r} being the particle's radius. Using our
model we calculate the dragging distance of the particle by treating
the pushing as a sequence of growing steps by the surface, each enabled
by the particle's diffusion away. Eventually the particle is engulfed
by ice growing around it when a rare event of long diffusion time
away from the surface occurs. By calculating numerically the statistics
of the diffusion times from the surface and therefore the probability
for a such a rare event we calculate the total dragging time and distance
\emph{L} of the particle by the ice front to be \emph{L}\textasciitilde{}
exp{[}1/(\emph{v\ensuremath{\centerdot}r}){]} where \emph{v} is the
freezing velocity. This relation for \emph{L} is confirmed by ours
and others experiments. The distance \emph{L} provides a length scale
for pattern formation during phase transition in colloidal suspensions,
such as ice lenses and lamellae structures by freeze casting. Data
from the literature for ice lenses thickness and lamellae spacing
during freeze casting agree with our prediction for the relation of
the distance \emph{L}. These results lead us to conjecture that lamellae
formation is dominated by their lateral growth which pushes and concentrates
the particles between them.
\end{abstract}
\maketitle

\section*{Introduction}

When a moving solidification front encounters a foreign particle in
the melt, it can either engulf it or push and reject it \citep{uhlmann1964interaction,asthana1993engulfment}.
The outcome of this interaction is fundamental in crystal growth of
single crystals \citep{ghezal2012observation,kulak2014one}, soil
freezing \citep{rempel2007formation}, alloy casting \citep{zhang2006inclusion}
and freeze casting \citep{wegst2010biomaterials,deville2008freeze,zhang2005aligned}.

Ice can only grow when there is water on its surface which then freeze.
Having a particle on its surface contradict this basic condition,
there are no water on the ice surface where the particle is and ice
can not grow there. The particle must be displaced from the surface
so its place on the ice surface would be replaced by water in order
for the ice to grow (fig \ref{fig:ice grow condition}). It is this
mechanism of moving the particle from the surface of ice that is in
the heart of the phenomenon of ice-particle interaction. 

For high front velocity and large particle size the particle will
be engulfed, as oppose to slow velocity and small particle for which
the particle can be pushed by the advancing crystallization front.
For a given particle size there exist a critical velocity \emph{v}\textsubscript{c}
that separates the two regimes.

Current models of this phenomenon rely on an actual force between
the particle and the ice surface which repels the particle. We name
these models as force balance models. Our model relies on the particle
being free to diffuse and no specific forces between the particle
and the surface are involved.

In the literature the interaction of ice surface and a particle is
modeled by a ``Force Balance Model''. In this model it is agreed
that there is a repelling force \emph{F\textsubscript{\emph{\textgreek{sv}}}}
between the ice surface and the particle which pushes the particle
ahead of the moving front \citep{asthana1993engulfment}. The pushing
force is calculated from Van der Waals interactions and interface
shape changes \citep{asthana1993engulfment}. This pushing force is
balanced by a drag force \emph{F\textsubscript{\emph{d}}} that pulls
the particle toward the solidification front \citep{rempel1999interaction}.
A steady state of constant velocity particle pushing is than reached
when the forces are equal $F_{d}=F_{\sigma}$. From this condition
the critical velocity can be estimated to scale as \footnote{The drag force in the liquid that opposes the particle movement is
$F_{d}\sim\eta vr^{2}$ , where \emph{\textgreek{h}} is the viscosity,
\emph{v} is the velocity of the particle and \emph{r} is the particle
radius. The disjointing pushing force is attributed to van der Waals
interaction between the particle, the ice surface and the thin liquid
film between them of thickness \emph{d} and to the surface energies
difference between them \textgreek{Dsv}\textsubscript{0} $F_{\sigma}\sim r\frac{\Delta\sigma_{0}}{d^{n}}$
where \emph{n} is a small integer (may be 1).} 
\begin{equation}
v_{c}\sim\frac{1}{\eta r}\label{eq:vc prop eta r}
\end{equation}

where \emph{r} is the particle radius and \emph{\textgreek{h}} is
the viscosity of the liquid. Equation \ref{eq:vc prop eta r} is the
widely accepted, experimentally verified and intuitive expression
for the critical velocity \emph{v}\textsubscript{c}\citep{uhlmann1964interaction,korber1985interaction,asthana1993engulfment,wegst2010biomaterials}.
We've omitted here the terms with the film thickness and the surface
energies because as we claim that the particle pushing is secondary
to a dominant phenomenon of the Brownian ratchet mechanism described
below, also they are almost inaccessible experimentally.

\section*{The Brownian Ratchet Model}

\subsection*{Outline of the Model}

\begin{figure}
\centering{}\subfloat[\label{fig:ice grow condition}]{\begin{centering}
\includegraphics[width=0.25\linewidth]{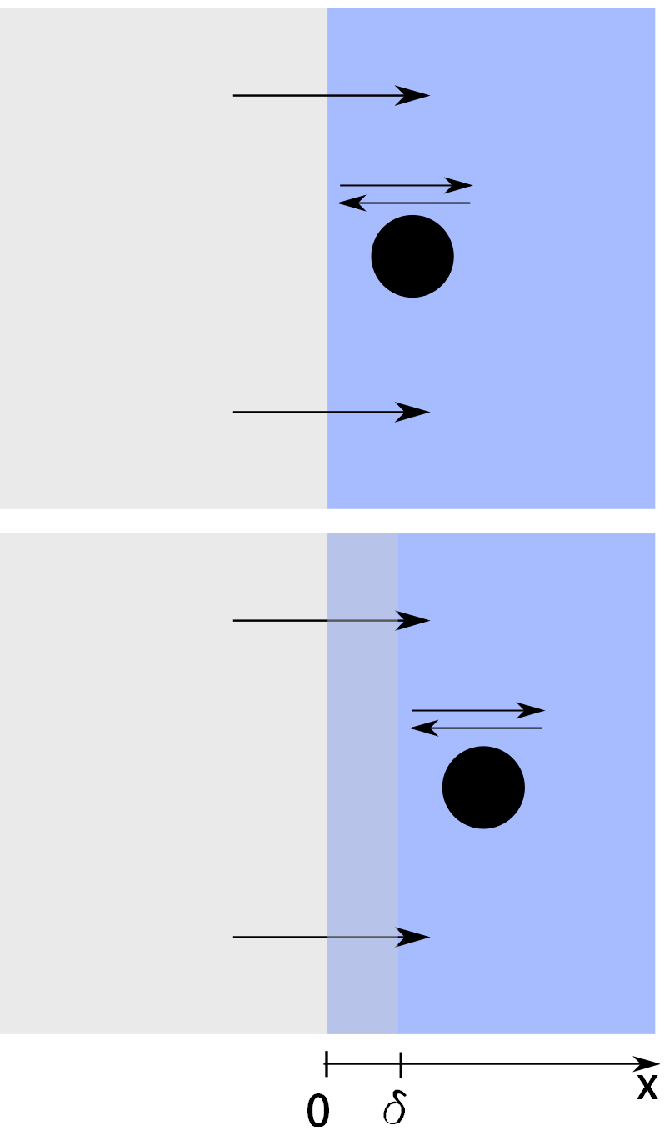}
\par\end{centering}

}\subfloat[\label{fig:simulation step growth}]{\centering{}\includegraphics[width=0.65\linewidth]{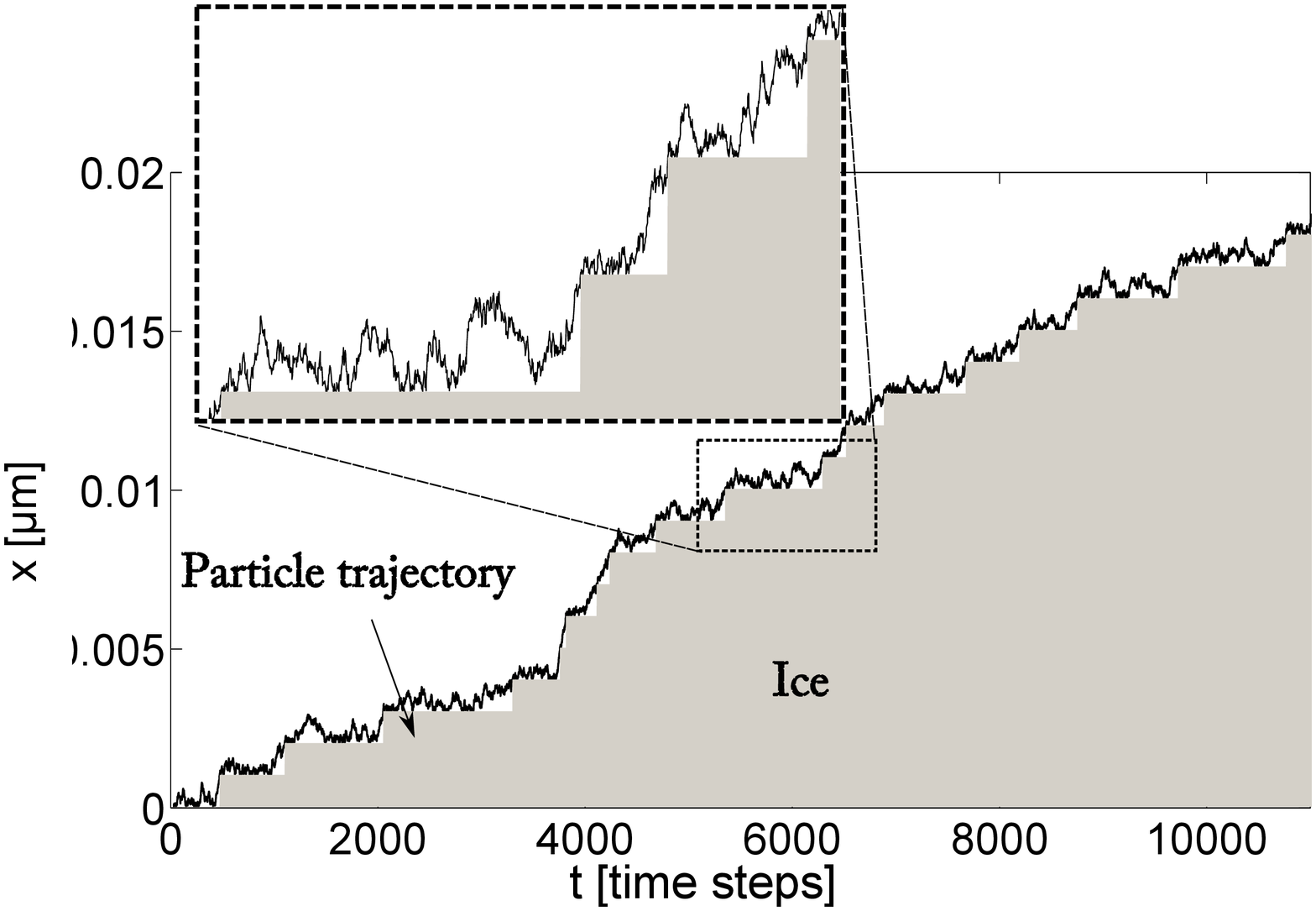}}\caption{\label{fig:Sketch-mechanism-simulation}Sketch and simulation of the
mechanism of Brownian ratchet. (a) The particle performs Brownian
motion and when the distance x between the particle and the ice front
exceeds \textgreek{d} (+thin film of constant width) another layer
of ice grows. (b) Simulation of the particle Brownian motion near
ice surface. The x-axes is the time of the simulation which is in
units of the number of time steps. A time step is $0.1\cdot\tau_{\delta}$.
Once the particle is over a distance \textgreek{d} from the ice surface,
ice grows a step.}
\end{figure}

We treat the particle as \textbf{free} to diffuse and to perform Brownian
motion in water near the ice surface. We approximate ice as a wall
that the particle can not penetrate. As long as the particle is close
to the ice surface, ice does not grow. Only when the particle is separated
at least a distance \emph{\textgreek{d} }from the surface can the
ice grow and than the particle is again on the surface (fig \ref{fig:Sketch-mechanism-simulation}).
There is no treatment in the literature of the case of Brownian motion
near ice surface. We justify the model's validity and assumptions
in a later section. 

\subsection*{The Critical Velocity}

\begin{figure}
\begin{centering}
\includegraphics[width=0.8\linewidth]{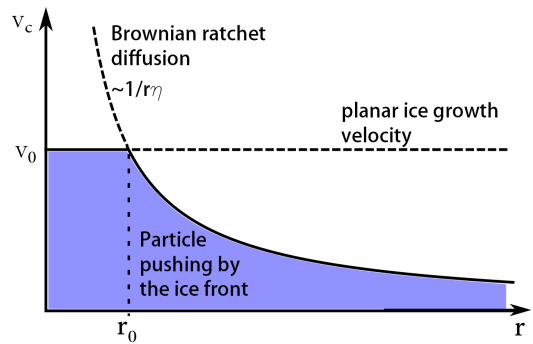}
\par\end{centering}

\caption{\label{fig: vc vs r}The planar growth velocity V\protect\textsubscript{0}is
limiting the particle pushing velocity for small particle sizes. For
large enough particle sizes the particle Brownian motion is limiting
the pushing velocity. }
\end{figure}

Now that we have formulated our model, we can calculate the velocity
of the particle motion according to the Brownian ratchet model by
\citet{peskin1993cellular}. The time to diffuse a distance \emph{\textgreek{d}}
is denoted as \emph{T}\textsubscript{\textgreek{d}}. A freely diffusing
particle travels a distance delta on average over time $\tau_{\delta}\equiv\frac{\delta^{2}}{2D}$.
For the most simple estimation \footnote{In more accurate model, the time to diffuse distance \textgreek{d}
should be the first passage time which is a little different than
our first simple estimation as the average time to diffuse. The first
passage time should scale as $<T_{\delta}>=b\tau_{\delta}=b\frac{\delta^{2}}{2D}$
where \emph{b} is a the first passage parameter and is dimensionless.
Using simulations (fig \ref{fig:Sketch-mechanism-simulation}) the
first passage parameter was estimated to $b=1.2$} $T_{\delta}=\tau_{\delta}$. After diffusing length \emph{\textgreek{d},}
another ice layer is added to the the ice front which means the particle
can not go back. Diffusing a larger distance $L=N\delta$ increases
the time linearly with \emph{N} $T_{L}=NT_{\delta}$ as appose to
$N^{2}$ for regular diffusion. The average speed of the particle
is than 
\begin{equation}
v=\frac{L}{T_{L}}=\frac{\delta}{T_{\delta}}=\frac{2D}{\delta}\label{eq:v perfect ratchet}
\end{equation}
 which is perfect ratchet velocity \citep{peskin1993cellular}. This
would be the speed if the particle would diffuse and ice would grow
instantaneously after it. Taking the diffusion coefficient from Einstein-Stokes
\citep{Einstein1905} relation with a correction $\xi$ \citep{brenner1961slow,pralle1998local}
for near wall diffusion $D_{particle}(r)=\frac{\xi kT}{6\pi\eta r}$
and inserting it into eq. \ref{eq:v perfect ratchet} we get the critical
velocity for engulfment 
\begin{equation}
v_{c}=\frac{\xi kT}{3\pi\eta r\delta}\label{eq:perfect ratchet full expresion}
\end{equation}
 which recovers relation (eq. \ref{eq:vc prop eta r}) which is well
established and agrees with the previous models. The derivation of
the expression for \emph{v}\textsubscript{c} in eq. \ref{eq:perfect ratchet full expresion}
is valid only for slow ice growth velocities because than the assumption
of instantaneous ice growth is not limited kinetically, i.e. the time
$\frac{\delta}{v}$ for ice to grow a distance \emph{\textgreek{d}}
is much longer than $\frac{\delta}{V_{0}}$ which is the fastest ice
can grow planar under specific conditions. Therefore, our model is
applicable for ice growth velocity $v\ll V_{0}$ and therefore for
large particles $r\gg r_{0}$ (fig \ref{fig: vc vs r}). Maximum planar
growth velocity of ice is estimated under typical conditions to $V_{0}\sim5\,\frac{\mu m}{s}$
\citet{hobbs2010ice}. Estimating the parameters of eq. \ref{eq:perfect ratchet full expresion}
with $kT=3.77\cdot10^{-21}J$, $\eta=1.67\cdot10^{-3}\frac{Js}{m^{3}}$,
$\xi=0.1$ yields 
\begin{equation}
v_{c}(r>15\,\mu m)\cong\frac{24}{\left(\unitfrac{r}{\mu m}\right)\cdot\left(\unitfrac{\delta}{nm}\right)}\frac{\mu m}{s}\label{eq:Vc big particles}
\end{equation}

The distance \emph{\textgreek{d}} is expected to be on the scale of
the size of an ice layer which is \textasciitilde{}0.5 \emph{nm}.
It is seen that for small particles the limiting factor in the critical
velocity for pushing the particle is the growth of ice. Therefore
the critical velocity does not depend on the particle size for small
particles (fig \ref{fig: vc vs r})
\begin{equation}
v_{c}(r<10\,\mu m)\cong V_{0}\label{eq:vc small particle const}
\end{equation}
This independence of the critical velocity on the particle size was
observed by \citet{uhlmann1964interaction} in one of the first papers
in the field. \citet{REMPEL2001} explained such a behavior by the
increased surface curvature of the ice surface near the particle for
small particles. In our model this behavior follows naturally from
the model.
This model has another interesting consequence. In the regime where
the critical velocity is constant (eq. \ref{eq:vc small particle const}),
V\textsubscript{0} is the maximum velocity ice can grow while staying
planar on a very small scale. This is a parameter that is hard to
measure optically (resolution limit) and to the best of our knowledge
was never measured.

\paragraph*{Affect Of The Temperature Gradient And The Particle Material }

Higher gradient stabilizes the surface \citep{glicksman2010principles},
so we expect for higher gradient the velocity V\textsubscript{0}
will be higher. Also the thermal gradient G and the thermal conductivities
of the liquid, crystal and particle $\kappa_{L},\,\kappa_{C},\,\kappa_{P}$
respectively are expected to affect \emph{$\delta=\delta(G,\kappa_{L},\kappa_{C},\kappa_{P})$
}. \citep{garvin2004drag,bolling1971theory}

\section*{Dragging time/distance of the particle by the ice front}

In freezing colloidal suspensions the particles are pushed by the
freezing front, concentrated and patterns appear. The scaling of this
pattern is important in freeze casting, inclusions in steel \citep{you2017modeling}
and for ice lenses formation.

\begin{figure}
\centering{}\includegraphics[width=0.9\linewidth]{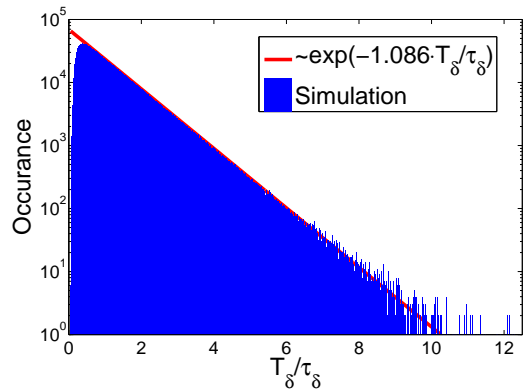}\caption{\label{fig:frequency distribution T}Frequency distribution of the
times \emph{T\protect\textsubscript{\emph{\textgreek{d}}}} as a function
of \emph{T\protect\textsubscript{\emph{\textgreek{d}}}}. The graph
summarizes the simulation of Brownian motion shown in figure \ref{fig:simulation step growth}.
Each bin is the number of times that the time to diffuse a distance
\emph{\textgreek{d} }was \emph{T\protect\textsubscript{\emph{\textgreek{d}}}}.
The red line is a fit to the data of equation \ref{eq:frequency distribution}
, where $N=5\cdot10^{6}$ was the number of times the simulation ran.}
\end{figure}

Pushing of the particle by ice front is limited in time (and in distance).
After being pushed by the ice for a time \emph{t} (or a distance \emph{L=v·t}),
the particle is being engulfed. The dragging of the particle is viewed
here as a sequence of growing steps by ice \{\emph{$T_{\delta}^{i}$}\}
(fig \ref{fig:simulation step growth}), each one enabled by the diffusion
away of the particle to distance\emph{ \textgreek{d}.} This pushing
of the particle is stopped once an event occurs that diffusing a \emph{\textgreek{d}}
takes time $T_{\delta}^{critical}\gg\frac{\delta}{v}$ and the particle
is engulfed. The total time of the particle dragging is 
\begin{equation}
t=\sum_{i=1}^{N}T_{\delta}^{i}=N<T_{\delta}>=N\frac{\delta}{v}\label{eq:drag time def}
\end{equation}
with N being the number of steps after which the particle engulfed,
i.e. $T_{\delta}^{N+1}=T_{\delta}^{critical}$. Next we'd like to
get the frequency distribution \emph{n}(\emph{T\textsubscript{\emph{\textgreek{d}}}})
of the times \emph{T\textsubscript{\emph{\textgreek{d}}}.} We preform
numerical simulation (fig \ref{fig:simulation step growth}) of a
Brownian motion first passage times with a reflecting boundary at
the origin (supplementary). The resulting distribution \emph{n}(\emph{T\textsubscript{\emph{\textgreek{d}}}})
of the times the particle reaches a distance \emph{\textgreek{d} }from
the origin for the first time is shown in figure \ref{fig:frequency distribution T}
and is best fitted as
 
\begin{equation}
n(T_{\delta})=\frac{N}{72}\exp(-1.08\frac{T_{\delta}}{\tau_{\delta}})\label{eq:frequency distribution}
\end{equation}

Engulfment of the particle once a critical event occurs
\begin{equation}
n(T_{\delta}^{critical})\simeq1\label{eq:critical occurrence}
\end{equation}
It is reasonable to assume that critical time is proportional to the
average diffusing time 
\begin{equation}
T_{\delta}^{critical}=A\frac{\delta}{v}\label{eq:critical time}
\end{equation}
, with $A\sim5-15$ being an empiric parameter.

Combining equations \ref{eq:drag time def}, \ref{eq:frequency distribution},
\ref{eq:critical occurrence}, \ref{eq:critical time} and the definition
of \emph{\textgreek{t}\textsubscript{\textgreek{d}}} we get 
\begin{equation}
1=\frac{t\cdot v}{72\,\delta}\exp(-\frac{1.08\,A\frac{\delta}{v}}{\frac{\delta^{2}}{2D}})\label{eq:critical occurance formula}
\end{equation}
than inserting the Einstein-Stokes relation to the diffusion coefficient
\emph{D,} the dragging time \emph{t} and distance \emph{$L=v\centerdot t$}
can be expressed as 
\begin{equation}
t=\frac{a\cdot\delta}{v}\exp(\frac{1.08\cdot2\cdot A\xi kT}{6\delta\pi\eta}\frac{1}{v\cdot r})\label{eq:dragging time}
\end{equation}
\begin{eqnarray}
L & = & v\cdot t=a\cdot\delta\exp(1.08\cdot2\frac{AD}{\delta}\frac{1}{v})=\label{eq:dragging distance}\\
 & = & a\cdot\delta\exp(\frac{1.08\cdot2\cdot A\xi kT}{6\,\delta\pi\eta}\frac{1}{v\cdot r})\nonumber 
\end{eqnarray}
where \emph{a} is a parameter we have replaced the numerical constant
with. For the case of the interface of water and ice, taking $T=273.15\,K$
, $\eta=0.018\frac{g}{cm\cdot s}$ than 
\begin{equation}
L=a\cdot\delta\exp(\frac{240\,A\xi}{\left(\unitfrac{r}{\mu m}\right)\cdot\left(\unitfrac{\delta}{nm}\right)\left(\nicefrac{v}{\frac{\mu m}{s}}\right)})\label{eq:dragging distance water-ice}
\end{equation}
This relation is verified in experiments where a single particle is
dragged by a growing ice surface with constant velocity (fig \ref{fig:dragging distance vs v}).
Our data and data from \citet{dedovets2017freezing} both agree with
equation \ref{eq:dragging distance water-ice}.

\subsection*{Multi particle systems}

This model explains the interaction between a single particle and
the interface of ice-water. Most interesting phenomenon relying on
this interaction are in systems with multiple particles interacting
with the ice surface, such as ice lens formation \citep{rempel2007formation}
and freeze casting \citep{deville2008freeze}. In such systems the
mutual diffusion coefficient is given by \citep{peppin2007morphological}
$D(\phi)=D_{0}\hat{D}(\phi)$ where $D_{0}$ is the Einstein-Stokes
diffusion coefficient and $\hat{D}(\phi)$ is a correction which depends
on the concentration $\phi$. Adjusting the diffusion coefficient
is the simplest correction to the model to account for the multi-particle
system. Instead of inserting $\hat{D}(\phi)$ as another parameter
to equations \ref{eq:dragging time} and \ref{eq:dragging distance}
the definition of \textgreek{x} can be altered, so it would be the
correction to diffusion which takes particle concentration into account.
Under this adjustment the particle dragging distances of multi-particle
system should be qualitatively the same as for single particle interacting
with the ice front. In figure \ref{fig:data-dragging-distance} (b-d)
the spacing in the resulting structure from colloidal solution is
plotted. The scaling $L\sim\exp(\frac{1}{v\cdot r})$ (eq. \ref{eq:dragging distance})
which was derived for single particle dragging is valid for multi-particle
systems such as ice lenses growth (fig. \ref{fig:ice lenses growth})
and for lamellae spacing (fig. \ref{fig:lamel spacing}).

\subsection*{Lamellae spacing }

\begin{figure}
\begin{centering}
\subfloat[\label{fig:dragging distance vs v}]{\centering{}\includegraphics[width=0.5\columnwidth]{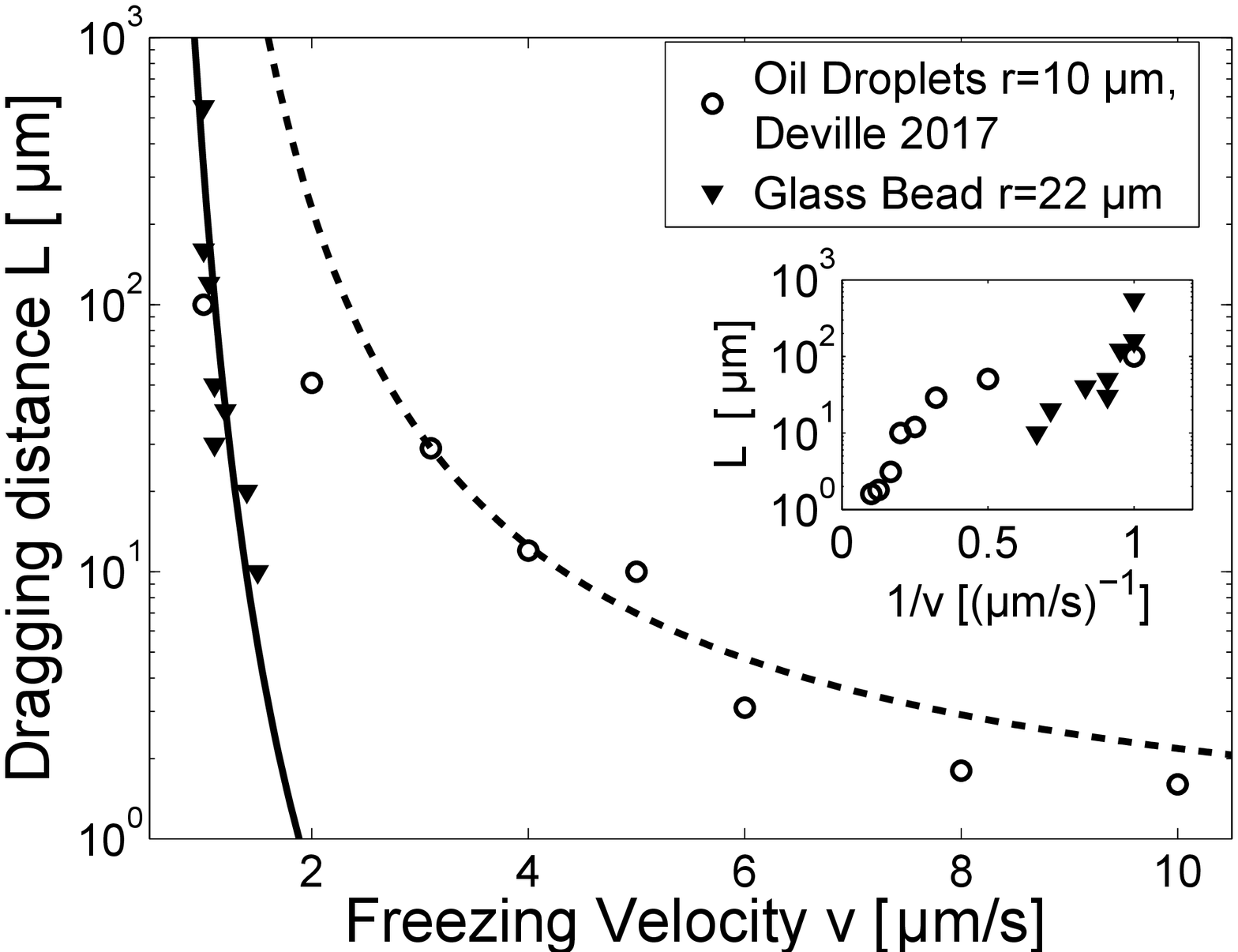}}\subfloat[\label{fig:ice lenses growth}]{\begin{centering}
\includegraphics[width=0.5\columnwidth]{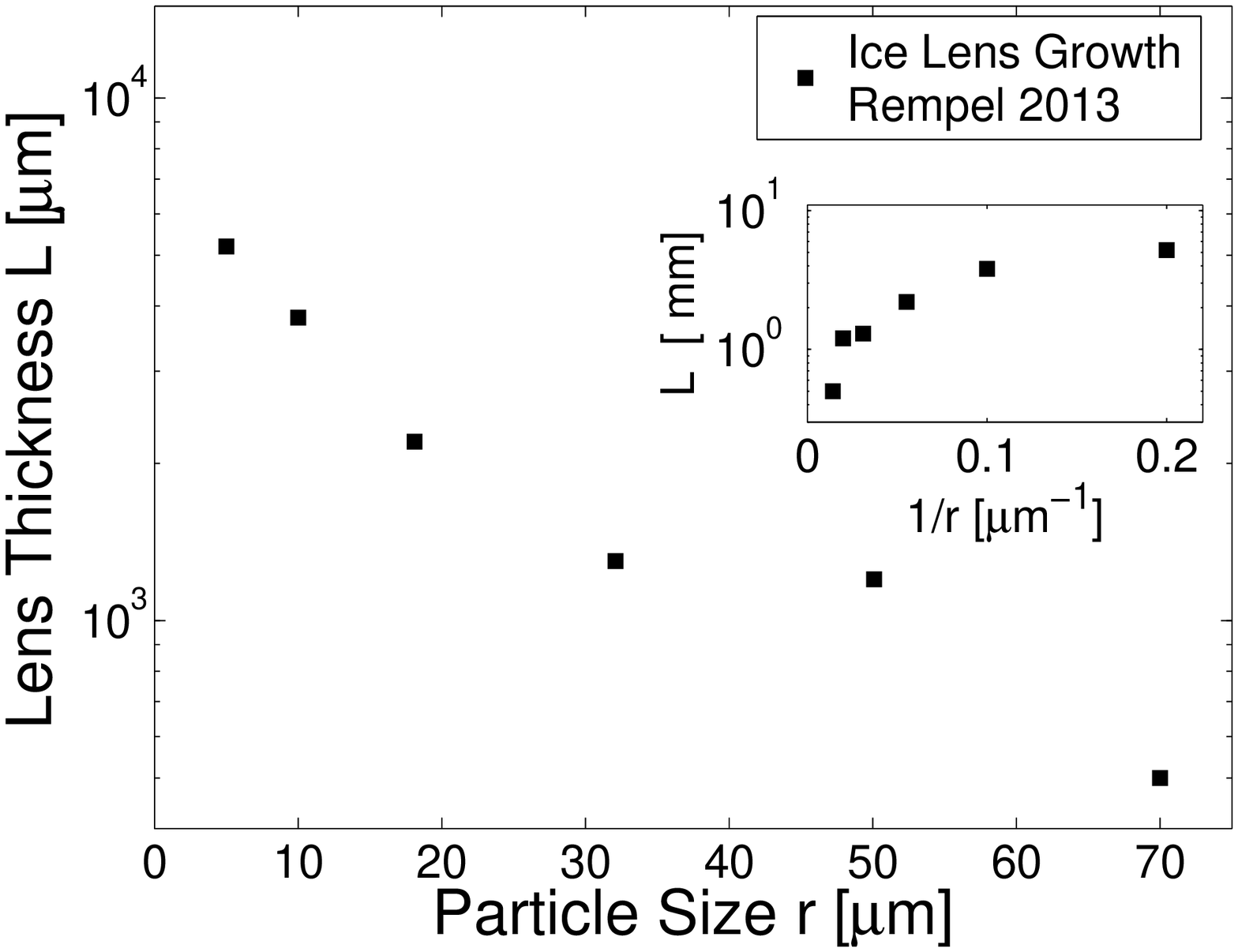}
\par\end{centering}

}
\par\end{centering}

\begin{centering}
\subfloat[\label{fig:lamel spacing}\citet{WASCHKIES2011}]{\begin{centering}
\includegraphics[width=0.5\linewidth]{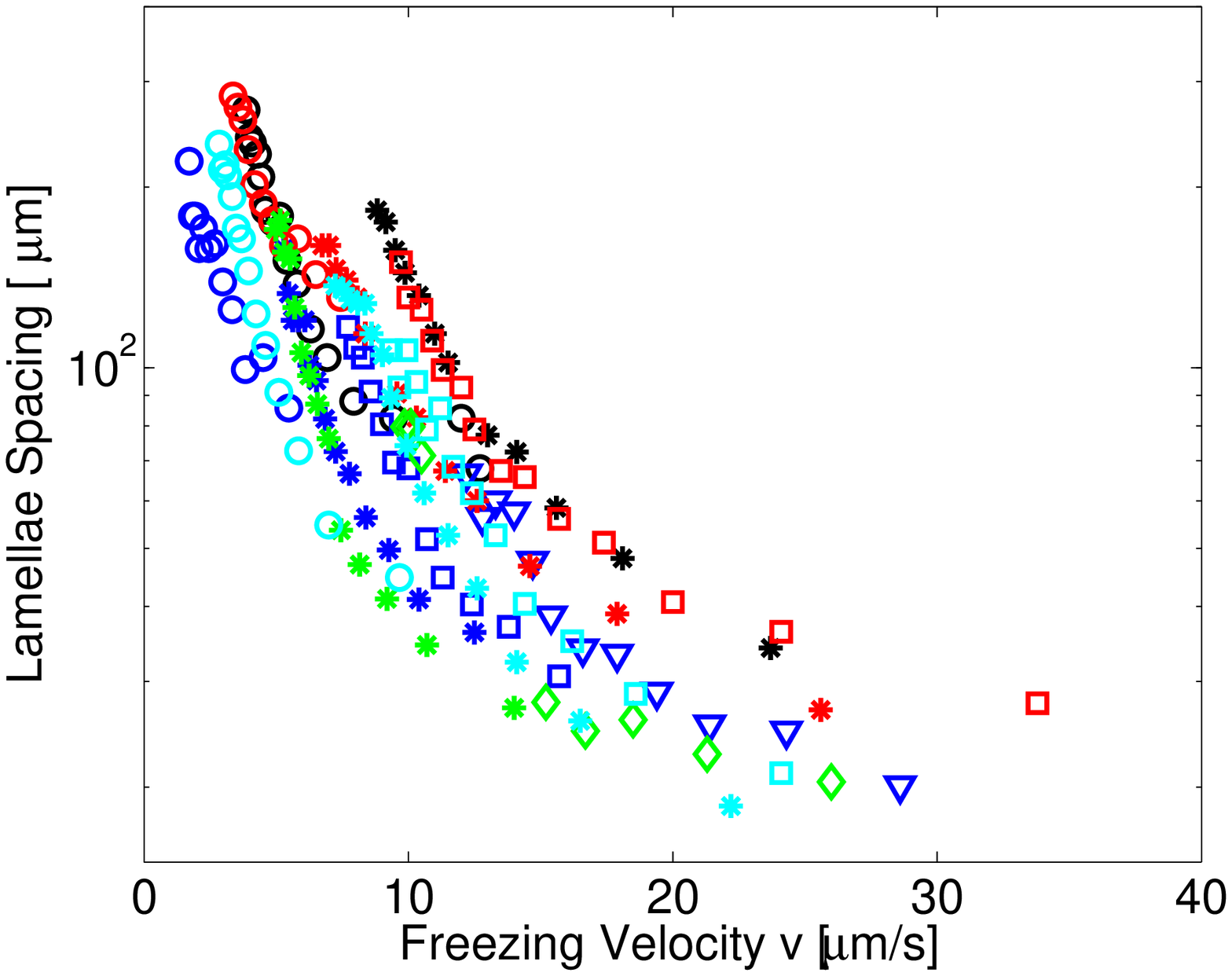}
\par\end{centering}

}\subfloat[\label{fig:lamel space recip}\citet{WASCHKIES2011}]{\begin{centering}
\includegraphics[width=0.5\linewidth]{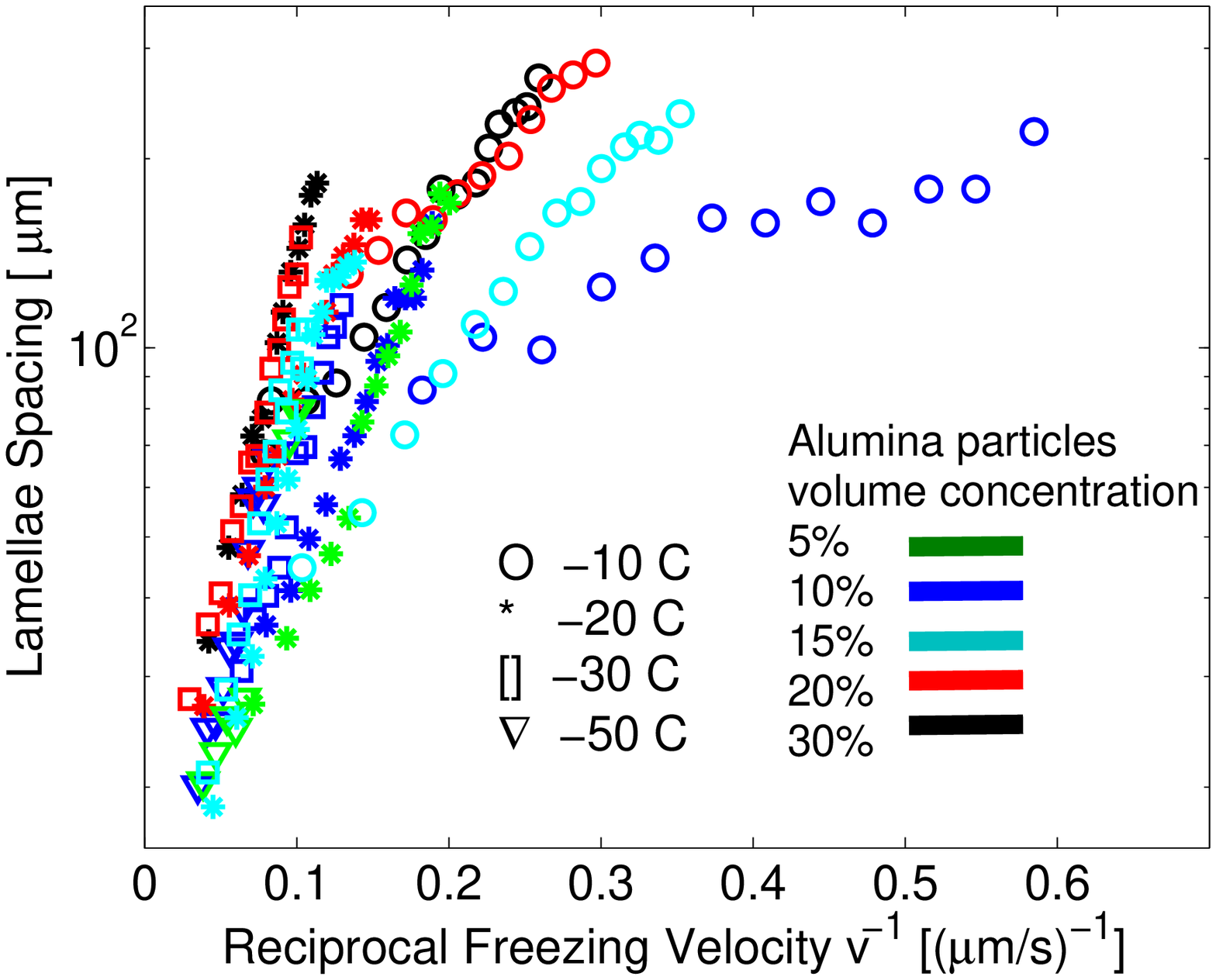}
\par\end{centering}

}
\par\end{centering}

\begin{centering}
\caption{\label{fig:data-dragging-distance}(a) The dragging distance of oil
droplets (\Circle ) and glass beads (\textifsymbol[ifgeo]{99}) for
different freezing velocities. The lines are fits of equation \ref{eq:dragging distance water-ice}
with \emph{r}= 10 and 22 \emph{\textgreek{m}m}, \emph{A\textgreek{x}}=0.59
and 0.17, \emph{\textgreek{d}}=1.9 and 0.2 \emph{nm}, \emph{n}=357
and 0.1 respectively. (b) (\textifsymbol[ifgeo]{80}) The thickness
of the ice lenses formed during directional freezing of a colloidal
solution of water and monodispersed glass beads. The y-axes are logarithmic
in all graphs. In the insets the x-axes is the reciprocal of the main
graph x-axes. A straight line in the inset corresponds to an agreement
with our model $L\sim\exp(\frac{1}{v\cdot r})$ (eq. \ref{eq:dragging distance}).
We see that the model (eq. \ref{eq:dragging distance}) describes
the data well. The data is taken from \citet{dedovets2017freezing}
for the oil droplets (\Circle ) and from \citet{saruya2013experimental}
for ice lenses growth (\textifsymbol[ifgeo]{80}). The data for the
glass beads (\textifsymbol[ifgeo]{99}) was measured using a standard
directional freezing setup for this study \citep{Directional}. (c-d)
The lamellar spacing as a function of the freezing velocity \emph{v}
(c) and $\frac{1}{v}$ (d). The structures formed during freeze casting
experiments for different freezing velocities of a solution of water
with 0.8 \emph{\textgreek{m}m} diameter alumina particles at different
volume concentrations in the of 5-30 \%. Different temperature of
the cooling plate were used -10, -20, -30 and -50 C, which represent
different thermal gradients at the interface. The data for the freeze
casting structures (c-d) is taken from \citet{WASCHKIES2011}.}

\par\end{centering}

\end{figure}

The scaling of the lamellae spacing (fig. \ref{fig:lamel spacing},
\ref{fig:lamel space recip} and supplementary) also obeys the relation
$L\sim\exp(\frac{1}{v})$. This is a surprising result, especially
since the agreement is so good. We interpret that as that the formation
of the spacing between lamellae is determined by the distance the
particles can be pushed between the lamellae. Lateral (with respect
to freezing direction) pushing of the particles between the lamellae
concentrates the solution. 

\subsection*{Justification of the model validity and assumptions}

Diffusion near a wall is highly damped \citep{brenner1961slow} due
to the friction (no-slip boundary condition) of the liquid near the
wall. The damping scales as $\frac{r}{d}$ depends on the thickness
\emph{d} of the liquid layer between the particle and the wall. \citep{uhlmann1964interaction,rempel1999interaction}
This dependence of the diffusion coefficient $\frac{\partial D}{\partial z}>0$
on the distance from the surface \emph{z} results in drift away from
the surface \citep{de2015hydrodynamics,lau2007state,PhysRevLett.99.138303}.
In our model we assume the distance between the particle and the ice
surface is roughly on the molecular scale and the variation in \emph{D}
can be neglected, so \emph{D} is assumed to be independent of \emph{z}. 

According to hydrodynamic calculations \citep{brenner1961slow,rempel1999interaction}
the diffusion coefficient of a particle near the ice surface should
vanish. Our model deals with distances between the particle and the
surface of ice being on the molecular level, it is unclear whether
the hydrodynamic approach is valid in this limit. The thing to notice
is that when the particle fluctuates away from the surface under pressure
is created between the particle and the surface which must be filled
with water molecules at the same rate that ice grows. The water molecules
can be supplied there by surface diffusion or by lubrication flow
\citep{rempel1999interaction} on the interface between ice and the
particle. The thickness of the liquid film between the particle and
the ice surface was considered to be from a molecular size \citep{uhlmann1964interaction}
up to 10 nm \citep{rempel1999interaction}. In both cases, for both
thicknesses it was considered that water can fill the gap between
the particle and the ice sufficiently fast so ice could grow in microns
per second.

Our model, as stated above, is the simplest approximation of the phenomenon
with the parameter $\xi$ being the correction term to the diffusion
taking these considerations into account.

\section*{Conclusions And Outlook}

The Brownian ratchet explains the mechanism behind the particle pushing
by the ice surface. We show here a \textbf{mechanical model} of how
ice is pushing particles. Using the model we have derived a relation
between the distance of the particle pushing and the particle size
and freezing velocity $L\sim\exp(\frac{1}{v\cdot r})$ (eq. \ref{eq:dragging distance}).
We showed that this scaling can be also used to estimate length scales
in patterns resulting from ice freezing and phase transition in general
of colloidal solutions. Using the data from\citet{WASCHKIES2011}
we saw that the lamellae spacing scales in the same way, which leads
us to conjecture that lamellae formation is driven by the pushing
of particles and not the surface instabilities on ice \citep{glicksman2010principles}.
We conjecture that the instability on the ice surface grows ahead
and starts to deform, loosing memory of its initial structure and
consequently a cellular structure emerges by pushing the particles
laterally between the cells.

It would be very interesting to see under which other conditions,
such as different particle concentrations in multi particle systems,
gravity and liquid flow relation \ref{eq:dragging distance} is still
valid. Obviously modifications to the model would be necessary to
properly describe and to account for these conditions. This model
provides a simple framework which can be further developed to describe
solution freezing. 

This model and this approach should be interesting for the problem
of membrane rupturing and consequently cell and tissue damaging by
growing ice. There ice grows until it approaches the membrane. Then
ice can no longer grow since there are no water available for freezing
between the ice and the membrane and we might naively think that the
cell is saved at this point. However, experiments show that ice can
grow into cells and rupture membranes under these conditions \citep{fuller2004life}.
A similar model to ours may be proposed where the thermal fluctuations
of the membrane play the same role as the particle's Brownian motion
in our model. The membrane fluctuations create a separation between
the membrane and the ice surface to allow ice to grow toward the membrane.
Once ice grows the membrane can return to its initial state and with
each such cycle it has a strain and tension build up that eventually
cause the membrane to rapture resulting in a catastrophe for the cell. 
\begin{acknowledgments}
M.C. acknowledges support from The Samuel and Lottie Rudin Scholarship
Foundation. M.C. would like to thank Prof. Stas Burov for useful discussions
during the summer school on stochastic processes with applications
to physics and biophysics in Acre.
\end{acknowledgments}

\bibliographystyle{aipnum4-1}
\bibliography{test2}

\end{document}